# Physical property characterization of *single step* synthesized NdFeAsO$_{0.80}$F$_{0.20}$ bulk 50K superconductor

V.P.S. Awana[1,*], R.S. Meena[1,2], Anand Pal[1], Arpita Vajpayee[1], K.V. R. Rao[2] and H. Kishan[1]

[1]National Physical Laboratory (CSIR), Dr. K.S. Krishnan Marg, New Delhi-110012, India
[2]Department of Physics, University of Rajasthan, Jaipur 302055, Rajasthan, India

We report an easy *single step* synthesis route of title compound NdFeAsO$_{0.80}$F$_{0.20}$ superconductor having bulk superconductivity below 50 K. The title compound is synthesized via solid-state reaction route by encapsulation in an evacuated (10$^{-3}$ Torr) quartz tube. Rietveld analysis of powder X-ray diffraction data shows that compound crystallized in tetragonal structure with space group *P*4/*nmm*. $R(T)H$ measurements showed superconductivity with $T_c$ ($R=0$) at 48 K and a very high upper critical field ($H_{c2}$) of up to 345 Tesla. Magnetic measurements exhibited bulk superconductivity in terms of diamagnetic onset below 50 K. The lower critical field ($H_{c1}$) is around 1000 Oe at 5 K. In normal state i.e., above 60 K, the compound exhibited purely paramagnetic behavior and thus ruling out the presence of any ordered FeO$_x$ impurity in the matrix. In specific heat measurements a jump is observed in the vicinity of superconducting transition ($T_c$) along with an upturn at below $T=4$ K due to the AFM ordering of Nd$^{+3}$ ions in the system. The Thermo-electric power (TEP) is negative down to $T_c$, thus indicating dominant carriers to be of n-type in NdFeAsO$_{0.80}$F$_{0.20}$ superconductor. The granularity of the bulk superconducting NdFeAsO$_{0.8}$F$_{0.2}$ sample is investigated and the intra and inter grain contributions have been individuated by looking at various amplitude and frequencies of the applied AC drive magnetic field.

*: Corresponding Author: e-mail: awana@mail.nplindia.ernet.in:
Fax/Phone – 0091-11-45609210/9310
Home Page: www.freewebs.com/vpsawana/

**INTRODUCTION**

Recent invention of superconductivity in Fe based compounds REFeAsO (RE=rare earth) [1] has taken the solid-state physics community by surprise. In fact after the avalanche of high $T_c$ superconductivity (*HTSc*) in 1986 [2] with superconducting transition temperature ($T_c$) of up to 130 K [3], although the invention of superconductivity in Borides viz. $MgB_2$ [4] came through as one of the major superconductors, but the Fe based compounds [1,5] really have been the biggest surprise. This is so because traditionally magnetic Fe/Fe based compounds were thought to be the last ones to exhibit superconductivity. The REFeAsO compounds as such are non superconducting and in fact magnetic (spin density wave) [6-8]. Host of these compounds were known decades before in literature and are recently reviewed by R. Pöttgen, D. Johrendt [9]. Although by now nearly two years are over after the discovery of superconductivity in $LaFeAsO_{1-x}F_x$ at 26 K [1], but active research on them is yet limited to few specialized laboratories. This is unlike the *HTSc* compounds. In fact the synthesis of REFeAsO is rather difficult in comparison to the *HTSc* compounds. Primarily because of the exothermic reaction of As while forming REAs or FeAs at around 500°C.

The ground state REFeAsO are charge compensated and not superconducting. To achieve superconductivity the carriers are introduced by $O^{2-}$ site $F^{1-}$ substitution or by introducing the $O_{1-x}$ oxygen vacancies, which in turn as a result of charge neutrality provide mobile carriers to the superconducting FeAs layer [1,6-8,10,11]. This is similar to that as in case of *HTSc*, where the other structural layers viz. BiO/TlO/HgO/Cu-O chains provide the mobile carriers to superconducting $Cu-O_2$ blocks. Though the O site F doping can be achieved via normal pressure solid state synthesis, for oxygen deficient superconducting samples extreme conditions of high pressure (up to 14 GPa) and high temperature (1200°C), i.e., *HPHT* are required [10,11]. *HPHT* is available at select laboratories and hence is not readily available to vast community of experimental condensed matter physicists/chemists. As far as the doping of F is concerned at O site, the reactivity of F with vacuum encapsulating material such as quartz at temperatures as high as 1150°C is a matter of concern. Basically, besides the rich physics of these compounds, the challenging task is to synthesize them, for which synergy between solid-state chemists and physicists are required. We synthesized the parent ground state LaFeAsO compound via a simple solid-state single step route, instead of multi step precursor methods [12] at normal pressure. In current letter we propose the same single step method to achieve superconductivity in NdFeAsO$_{0.80}$F$_{0.20}$. In fact we were prompted to do so following a very recent article [13], where the process time of as short as couple of hours is proposed for fine particle ball milled SmFeAsO/F superconductors. However in this article [13], the precursors SmAs and FeAs are used unlike as in our case, where we start with pure Fe and As and in a *"single step"* the final compound is achieved. The resultant NdFeAsO$_{0.80}$F$_{0.20}$ compound is a near single phase bulk superconductor with superconductivity of up to as high as 50 K, which is competitive or even better than many reported results on similar compounds being processed with multi steps and rather lengthier heating schedules. Host of physical properties including structural details, transport (electrical, thermal) and magnetic under applied fields of up to 14 Tesla are reported here for NdFeAsO$_{0.80}$F$_{0.20}$ superconductor.

**EXPERIMENTAL DETAILS**

The title compound was synthesized via solid-state reaction route. Stoichiometric amounts of As, Fe, NdF$_3$, Nd, and Fe$_2$O$_3$ of better than 3N purity were mixed in ratio of 1/15(NdF$_3$)+14/15(Nd)+4/15(Fe$_2$O$_3$)+7/15(Fe)+As in a glove box of <1ppm level of Oxygen and humidity atmosphere. The mixed powder was pelletized in rectangular bar shape and further encapsulated in an evacuated (10$^{-3}$ Torr) quartz tube. Sealed and evacuated quartz tube was heat treated for 12 hours each at 550$^o$C, 850$^o$C and 1150$^o$C. Finally the sample was allowed to cool down to room temperature. All the heating schedules were given in one single step. The resultant compound was hard enough for transport measurements and was black in color. Typically, 1 gram of raw pellet was sealed in 2.5 cm diameter and 10 cm length high quality "*thick walled*" quartz tube. The X-ray diffraction pattern of the compound was taken on a Rigaku diffractometer using CuK$_\alpha$ radiation and the Rietveld analysis was carried out to know the lattice parameters and impurities etc., if present at all. Magneto-resistivity measurement *R(T)H* up to 14 T Field and *DC* magnetic susceptibility in both zero-field-cooled (*zfc*) and field-cooled (*fc*) situations along with the isothermal magnetization (*MH*) measurements were carried out on a physical property measurement system (*PPMS*) from Quantum Design (*QD*). The AC susceptibility at different frequencies (33 to 9999 Hz) & different amplitudes (1 to 15 Oe) of AC drive field with temperatures down to 5 K, the specific heat with temperature down to 2.2 K and thermo-electric power (TEP) were also measured on the same physical property measurement system (*PPMS*).

**RESULTS AND DISCUSSION**

Figure 1 depicts the room temperature X-ray diffraction (*XRD*) patterns of NdFeAsO and NdFeAsO$_{0.80}$F$_{0.20}$. The studied sample is crystallized in tetragonal structure with *P*4/*nmm* space group. The lattice parameters are $a$ = 3.969(1) Å and $c$ = 8.596(3) Å for NdFeAsO and $a$ = 3.9561(6) Å and $c$ = 8.539(2) Å NdFeAsO$_{0.80}$F$_{0.20}$. The decrease in *c*-parameter and the volume is indicative of successful substitution of F at O site. These results are in agreement with earlier reports on similar compounds being synthesized by various other routes [14,15]. Generally speaking the studied compounds are phase pure with only minute amount of some unidentified species for F doped sample, which are marked on XRD pattern with * in Figure 1. Rietveld analysis is carried out in space group *P*4/*nmm* with Wyckoff positions as Nd(2c), O/F(2a), Fe(2b) and As(2c). The Rietveld analysis fitted positions within mentioned scheme of the co-ordinates are Nd (¼, ¼, z), Fe (¾, ¼, ½), As (¼ ¼, z) and O/F (¾, ¼, 0) give Nd (*z*) at 0.138 and 0.151 and the As (z) at 0.658 and 0.6514 respectively for NdFeAsO and NdFeAsO$_{0.80}$F$_{0.20}$ samples. The fit parameters obtained are $R_p$ = 4.16, $R_{wp}$= 5.34, $R_{exp}$= 3.46 and $\chi^2$= 2.37 for NdFeAsO and $R_p$ = 5.67, $R_{wp}$= 7.52, $R_{exp}$= 3.56 and $\chi^2$= 4.46 for NdFeAsO$_{0.80}$F$_{0.20}$ sample. Obviously the fit parameters are slightly better for NdFeAsO as compared to NdFeAsO$_{0.80}$F$_{0.20}$ sample, because the later contains small-unidentified impurities. In general, the variation of lattice parameters and coordinate positions along with the fit parameters are in agreement with previous reports on similar compounds [14,15].

Resistance versus temperature (*R-T*) plot for the single step normal pressure synthesized NdFeAsO$_{0.80}$F$_{0.20}$ compound is shown in Figure 2 (a). The *R-T* of ground state NdFeAsO compound is shown in inset of Figure 2 (a), which is not superconducting and rather exhibits a spin density wave (SDW) characteristic step in resistivity below around 160 K. This is in confirmation with other reports on similar compounds [6-8,14,15]. For NdFeAsO$_{0.80}$F$_{0.20}$ compound, the room temperature resistance is around 3.2 mΩ, which decreases to 0.6 mΩ before superconducting onset at around 51 K. The resistivity is decreased by over a factor of six before the superconductivity sets in at below 51 K, thus exhibiting good normal state metallic behavior of the single step normal pressure synthesized NdFeAsO$_{0.80}$F$_{0.20}$ sample. Though superconductivity sets in at $T_c$ (onset) of 51 K, the superconductivity is finally established at $T_c$ ($R$=0) of 48 K.

The normal state $R(T)$ behavior shows linear dependence on temperature. In Figure 2(a) the blue solid line shows the fitted resistance plot according to equation $R = R_0 + AT$, where $R_0$ is the residual resistance and A is the slope of the graph. Interestingly the experimental data (red dots) for the entire temperature range in the normal state are well fitted according to this linear equation. The values of $R_0$ and A found as to be 3.22E-8 $\Omega$ and 4.03E-8 $\Omega$/K respectively. This behavior is quite different from any other superconductor like $MgB_2$, HTSc and even from some of the iron-pnictides. Bhoi et al. found that $\rho$ exhibits a quadratic temperature dependence, $\rho = \rho(0) + AT^2$, in the temperature range 70 K $\leq T \leq$ 150 K, while it is linear in $T$ in the intermediate region, 170 K$\leq T \leq$ 270 K for $PrFeAsO_{0.6}F_{0.12}$ [16,19]. For $MgB_2$ superconductor normal state resistivity in the high temperature range shows the linear temperature dependence due to the electron phonon interaction. According to the solid-state theory, metals exhibit linear temperature dependence of resistivity when $T > \theta_D/5 \sim \theta_D/4$, where $\theta_D$ is the Debye temperature. Further, in the range from 40 to 150 K resistivity curve is well fitted by a power law $\rho(T) = \rho_0 +$ const x $T^m$ with $m$=3-5, where $\rho_0$ is the residual resistivity. This behavior is well established in $MgB_2$ [20-23]. For HTSc compounds, at the hole concentration for optimum $T_c$, the in-plane resistivity ($\rho_{ab}$) is linear in temperature over a wide temperature range from just above $T_c$ to nearly 1000 K [24,25]; but before the superconductivity occurs resistivity deviates from its linear behavior and $\rho(T)$ curve gets some rounding and then resistance drops abruptly to nearly zero. This non-linearity was explained by the occurrence of pseudo-gap [26-29]. The temperature below which the resistivity apparently deviates from the $T$-linear behavior is called $T_s$ and is thought to correspond to opening of the pseudo-gap [26-29].

Figure 2(b) shows the temperature dependence of the electrical resistance $R(T)H$ in applied magnetic field from 0 to 14 Tesla. The resistive transition for this superconducting compound shifts to lower temperature by applying the magnetic field [Figure 2(b)]. The transition width becomes wider with increasing strength of $H$, a characteristic of type-II superconductivity. Here, we define a transition temperature $T_c(H)$, which satisfies the condition that $\rho(T_c,H)$ equals to 90% of the normal-state value ($\rho_N$) for applied field $H$. The $T_c(H)$ values are 50.65 K and 48.94 K for 0 and 14 Tesla field respectively. The calculated value of $dH_{c2}/dT$ is thus 8.2 Tesla/Kelvin. The transition width $\Delta T_c = T_c(90\%) - T_c(10\%)$ are 2.98 and 9.82 K for 0 and 14 Tesla respectively. For LaFeAsO$_{0.89}$F$_{0.11}$ and PrFeAsO$_{0.6}$F$_{0.12}$ the value of $\Delta T_c$ is 4.5 K and 2.7 respectively at zero field [17, 30]. Thus the superconducting transition for present NdFeAsO$_{0.80}$F$_{0.20}$ sample is sharper than that for LaFeAsO$_{1-x}$F$_x$ [1,16] and nearly same to that for PrFeAsO$_{0.6}$F$_{0.12}$ [17].

The upper critical field is determined using different criterion of $H_{c2}=H$ at which $\rho=90\%\rho_N$ or $50\%\rho_N$ or $10\%\rho_N$, where $\rho_N$ is the normal resistivity or resistance at about 51 K. The $H_{c2}$ variation with temperature is shown in Figure 2(c). To determine $H_{c2}(0)$ value, we applied Ginzburg Landau (GL) theory. The GL equation is:

$$H_{c2}(T)=H_{c2}(0)*(1-t^2)/(1+t^2)$$

Where, $t=T/T_c$ is the reduced temperature [18]. The fitting of experimental data is done according to the above equation, which not only determines the $H_{c2}$ value at zero Kelvin [$H_{c2}(0)$] but also determines the temperature dependence of critical field for the whole temperature range. $H_{c2}(10\%)$, $H_{c2}(50\%)$ and $H_{c2}(90\%)$ are estimated to be 62, 115 and 345 Tesla respectively at 0 K.

The corresponding $H_{c2}(0)$ value derived from the Werthamer–Helfand–Hohenberg (WHH) formula is $0.693T_c|dH_{c2}/dT|_{Tc} \approx 285$ T. Considering that the WHH evaluation underestimates $H_{c2}$ in two-band superconductors [19], huge upper critical fields are expected in these compounds. More certain evaluations will require magneto-resistivity measurements at much higher applied magnetic field.

*DC* magnetic susceptibility versus temperature plot in both zero-field-cooled (*zfc*) and field-cooled (*fc*) situations is given in Figure 3. The applied magnetic field is 50 Oe. Superconductivity sets in below 49 K, as evidenced from clear diamagnetic signal at this temperature. Although the transition seems slightly broad the same is still sharper than previous reports on polycrystalline NdFeAsO/F [14,15]. Both *zfc* and *fc* are nearly saturated below 30 K. As far as the determination of superconducting volume fraction is concerned the same cannot be ascertained without ambiguity primarily due to positive paramagnetic contribution to magnetic susceptibility from Nd/Fe moments and secondly the pinning defects and impurities hamper the real outcome. Still one can say with confidence that the studied compound is a bulk superconductor with shielding fraction above 25% and superconducting volume fraction of > 8%.

Measurement of the AC susceptibility can be conveniently used to investigate both intergrain and intragrain vortex dynamics in superconductors [31-33]. Figure 4 represents the plot of real ($M'$) and imaginary ($M''$) components of AC susceptibility as a function of temperature obtained at the frequencies 33, 333, 999, 3333, 6666 & 9999 Hz at AC field amplitude of 10 Oe for NdFeAsO$_{0.80}$F$_{0.20}$ sample. It is noticed that the near $T_c$ region is almost independent of value of frequency i.e. the diamagnetic onset transition temperature is the same for all the frequencies. The $M'(T)$ curves show the typical two-step transition due to intergranular (at low $T$) and intragranular (near $T_c$) response [31-33]. On the other hand, a well-defined intergranular peak (in correspondence to the low-$T$ step) is observed for all the frequencies in the imaginary component of susceptibility ($M''$) at a characteristic temperature $T_p$. This peak arises when the AC field penetrates through the intergranular region just to the center of the sample. On increasing the frequency, the peak intensity increases and $T_p$ shift toward higher temperatures. $T_P$ varies from 18.7 to 34.7 K, when the frequency is increased from 33 to 9999 Hz. Bonsignore et. al [33] investigated the frequency dependence of AC susceptibility in LaFeAsO$_{0.94}$F$_{0.06}$ and observed a $T_p$ shift of about 4 K varying the frequency from 10 kHz to 100 kHz. The results are comparable and clearly demonstrate the fact that studied bulk polycrystalline NdFeAsO$_{0.80}$F$_{0.20}$ sample is a granular superconductor like other reported oxy-pnictides.

The temperature variation of the AC susceptibility measured at different AC field amplitudes 1, 2, 4, 6, 8, 10, 12, 13 & 15 Oe is shown in Figure 5. The diamagnetic onset temperature is ~ 47.7 K at 1 Oe and it is almost constant at different fields. At highest field of 15 Oe the $M''$ data show both the intergranular (at low $T$) and intragranular (near $T_c$) peaks however at lower fields only intergranular peak is observed. The kink at ~ 45 K for higher amplitudes is reminiscent of intra-grain superconductivity. This means at these amplitudes the inter-grain superconductivity due to decoupling of individual grains is nearly disappeared and that of intra-grain (individual grain) is appeared. The intergranular peak temperature $T_p$ decreased from 38.68 K to 5.69 K, when the applied field amplitude is increased from 1 Oe to 15 Oe. This in agreement with Bonsignore et. al [33]; they also got the decrement in the $T_p$ values with increasing the AC field amplitudes. Bonsignore et. al noticed a strong field dependence even at low fields (~0.007 Oe); it is worth to remark that in high $T_c$ cuprates $T_p$ nearly saturates for $H_{ac}$<0.1 Oe [34,35]. It is also noted that values of maximum $M''(T_p)$ is increasing monotonically with increasing field. We calculated the intergranular critical current density ($J_c$) in the present sample using Bean's model from the relation $J_c(T_p) = H_a/a$ for a sample having cross section 2a X 2b where a<b [36,37]. $J_c(T_p)$ is the intergranular critical current density at $T_p$ -the temperature of the $M''$ peak, $H_a$ is the amplitude of the applied AC field. Inset of Fig.5 shows the variation of $J_c$ with respect to $T_p$.

Figure 6 shows the isothermal magnetization (*MH*) loops of the studied NdFeAsO$_{0.80}$F$_{0.20}$ compound at 2 and 5 K, with applied fields of up to 15 kOe. The *MH* loops are wide open in superconducting state and thus warrants again the establishment of bulk superconductivity in the studied sample. The lower critical field ($H_{c1}$) as being viewed from inversion point of the *MH* loops is around 1000 Oe at 5 K. As far as upper critical field ($H_{c2}$) is concerned the *MH* loops are wide open for higher fields of up to 40 kOe as well (see inset Figure 6), this is in agreement with ref. 15. In fact like *HTSc* cuprates the upper critical field of FeAs based oxypnictide superconductor is known to be very high, and their *MH* loops remain open for very high applied fields. We estimated the critical current ($J_c$) of studied NdFeAsO$_{0.80}$F$_{0.20}$ compound (3.54mm x 1.53mm x 7.34mm) from Bean's critical state model at 5 K under one Tesla field. The $J_c$ (1 Tesla, 5 K) comes out to be 2.6x10$^4$ A/cm$^2$, which is comparable to that or better than as reported in ref. 14 for similar bulk samples.

One of the main culprits hampering the superconductivity of REFeAsO based superconductors is the presence of un-reacted pure Fe or FeO$_x$ in the superconducting matrix. Both metallic/oxidized irons (FeO$_x$) are known to order magnetically above or close to room temperature. This has already been highlighted in literature [38,39]. One of the ways to rule out the presence of minute impurity of ordered FeO$_x$ in ReFeAsO is either Mössbauer spectroscopy [38] or the magnetization in normal state i.e., between superconducting transition and room temperature (300 K). To rule out the presence of magnetically ordered FeO$_x$ we carried out the magnetization measurements i.e. both *M*(*T*) and *M*(*H*) between 60 K (above superconducting transition) and 300 K. The *M*(*T*) of presently studied NdFeAsO$_{0.80}$F$_{0.20}$ compound is shown in Figure 7. The measurements are carried out between 60 K and 300 K under applied field of 10 kOe. As evidenced from this figure the studied NdFeAsO$_{0.80}$F$_{0.20}$ compound is paramagnetic in nature without any traces of ordered Fe impurity. The $\chi$ (*T*) values are close to that as reported earlier [14,15]. The magnetic susceptibility $\chi$ (*T*) follows the Curie-Weiss paramagnetic behavior and the $\chi^{-1}$ (*T*) behavior is purely linear (see Figure 7). The fitting of $\chi$ (*T*) data gives the total paramagnetic moment close to 3$\mu_B$. In studied NdFeAsO$_{0.80}$F$_{0.20}$ both Nd and Fe are in paramagnetic state. The Fe moment in REFeAsO is close to 0.5 to 0.7$\mu_B$ [40,41]. On the other hand $Nd^{4f}$ moment is itself close to 3$\mu_B$. Infact the moment values deduced from magnetization measurements are only the average values and as such are screened due to crystal field effects and hence we do not stress upon exact numerical values of the Nd/Fe moments. Albeit, our $\chi(T)$ behavior and numerical values are close to the ones as reported in ref. 15. The $\chi^{-1}(T)$ plot intersects the y-axis in negative directions, indicating the possibility of anti-ferromagnetic ordering of Nd moments at low temperature. Further to rule out the presence of any ordered foreign $FeO_x$ phase the $M(H)$ is carried out at 200 K in varying fields of up to 3kOe, see inset Figure 7. The linearity of $M(H)$ at 200 K excludes the possibility of ordered foreign impurities in the studied "*single step*" synthesized superconducting $NdFeAsO_{0.80}F_{0.20}$ compound.

In Figure 8, we show the specific heat $C_p(T)$ of $NdFeAsO_{0.80}F_{0.20}$ in the temperature range of 2.2 to 200 K at zero field. The observed value of $C_p$ at 200 K is around 95 J /mol K. With decrease in $T$ the $C_P$ goes down continuously and an apparent feature in $C_p$ results in the vicinity of the $T_c$ corresponding to superconducting transition. To have a clear view of this anomaly, we plotted heat capacity as $C_p/T$ versus $T^2$ in the lower inset of Figure 8. The kink is clearly seen and is related to the formation of superconducting state. The bulk nature of superconductivity is thus confirmed in the studied sample by the present $C_p$ measurement.

Upper inset of Figure 8 shows the $C_p$ vs $T$ plot at lower temperature. Below 4 K an upturn is observed in $C_p$ plot; this upturn is due to the antiferromagnetic ordering of Nd at low temperature. Since we could perform the measurement till 2.2 K only therefore we did not get the complete peak otherwise AFM ordering of Nd ions occurs below 1.8 K. For LaFeAsO without the rare earth 4f electrons, there is no such specific heat anomaly at low temperature [42] indicating that the AFM transition for this title compound is originated from the ordering of Nd 4f moments. Chen et. al also observed the AFM ordering of the Ce 4f electrons at 3.7 K under zero magnetic field in CeOFeAs compound [43]. In SmFeAsO compound this anomaly at low temperature is also observed by some of us at around 4.5 K due to the AFM ordering of $Sm^{3+}$ spins [44]. Considering the decreasing magnetic moments of trivalent Sm, Ce and Nd, the AFM ordering temperatures of 5.4, 3.7 & below 2.2 K respectively seems reasonable.

Since the bulk superconductivity was realized by fluorine doping, the electron doping is expected by the introduction of oxygen deficiency in F-doped $NdFeAsO_{0.80}F_{0.20}$ sample. In order to confirm this expectation and to provide the direct evidence, the thermo-electric power (TEP) is measured. Figure 9 shows the temperature dependence of thermo-electric power $S(T)$ for NdFeAsO$_{0.80}$F$_{0.20}$ sample. This compound shows negative TEP over the entire temperature range, it has a positive slope and changes almost linearly with $T$ i.e. with decreasing temperature negative TEP increases. S varies from -26 µV/K at 250 K to a value of -48 µV/K at ~110K then decrease in magnitude as the temperature is lowered further. At superconducting transition temperature ~50 K, TEP sharply drops to zero. The maximum in the data is probably due to the competition between dominant electron like bands and the expected proximity of hole-like bands near the Fermi energy. Negative TEP indicates that the dominant carriers are electron in the superconducting NdFeAsO$_{0.80}$F$_{0.20}$ system. A qualitatively same type of the TEP behavior has been reported for SmFeAsO$_{0.80}$F$_{0.20}$ in Ref. 45. In the parent compound NdFeAsO, a high temperature part of the $S(T)$ dependences look similar to the F-doped sample, while at low temperatures the thermopower of parent compound (NdAsFeO) develops a broad maximum below $T \sim 200$ K [45]. This maximum appears due to formation of the SDW state, since such a behavior has been already described theoretically [46].

Summarily we have been able to synthesize bulk superconducting NdFeAsO$_{0.80}$F$_{0.20}$ compound with superconductivity transition temperature as high as 50 K via a simple "*single step*" solid-state reaction route. The single step route minimized the synthesis time and at the same time produced high quality bulk superconducting compound. Host of physical properties including structural details, transport (electrical, thermal) and magnetic under applied fields of up to 14 Tesla are reported here for NdFeAsO$_{0.80}$F$_{0.20}$ superconductor.

**ACKNOWLEDGEMENT**

Anand Pal and Arpita Vajpayee would like to thank the *CSIR* for the award of Senior Research Fellowship to pursue their *Ph. D* degree. Authors thank Prof. R.C. Budhani, DNPL for his keen interest and encouragement for superconductivity research.

**FIGURE CAPTIONS**

Figure 1: The observed (red dots), calculated (solid line) and differences diffraction (bottom solid line) profiles at 300 K for NdFeAsO and NdFeAsO$_{0.80}$F$_{0.20}$ samples.

Figure 2 (a): Temperature dependence of the resistance $R(T)$ of NdFeAsO$_{0.80}$F$_{0.20}$ sample and inset shows $R(T)$ of the NdFeAsO sample.

Figure 2 (b) & (c): The transition zone of $R(T)H$ plot of NdFeAsO$_{0.80}$F$_{0.20}$ sample in applied fields from 0 to 14 T and the $H_{c2}$ vs $T$ plots derived from $R(T)H$ measurements using GL equation.

Figure 3: $M(T)$ of NdFeAsO$_{0.80}$F$_{0.20}$ sample in both zfc and fc situation under applied field of 50Oe.

Figure 4: Plot of real ($M'$) and imaginary ($M''$) components of AC susceptibility as a function of temperature at the frequencies 33, 333, 999, 3333, 6666 & 9999 Hz for fixed amplitude at 10 Oe of NdFeAsO$_{0.80}$F$_{0.20}$ sample. Arrows indicate the increasing frequency values.

Figure 5: Plot of real ($M'$) and imaginary ($M''$) components of the AC susceptibility versus temperature, measured in the NdFeAsO$_{0.80}$F$_{0.20}$ sample at the AC field amplitudes 1, 2, 4, 6, 8, 10, 12, 13 & 15 Oe at a fixed frequency 333 Hz and inset shows the $J_c$ versus $T_p$ behavior for the same sample.

Figure 6: $M(H)$ of NdFeAsO$_{0.80}$F$_{0.20}$ sample at 2 and 5K, inset shows the same at 4K for higher fields.

Figure 7: $M(T)$ of NdFeAsO$_{0.80}$F$_{0.20}$ normal state sample, inset shows the $M(H)$ at 200K for the same.

Figure 8: Specific heat of NdFeAsO$_{0.80}$F$_{0.20}$ sample (main panel). Lower inset shows the specific heat as $C_p/T$ vs $T^2$ and upper inset shows the enlarged view of specific heat of NdFeAsO$_{0.80}$F$_{0.20}$ in the low temperature range (below 20 K).

Figure 9: Temperature dependence of thermo-electric power $S(T)$ for the sample NdFeAsO$_{0.80}$F$_{0.20}$.

Figure 1:

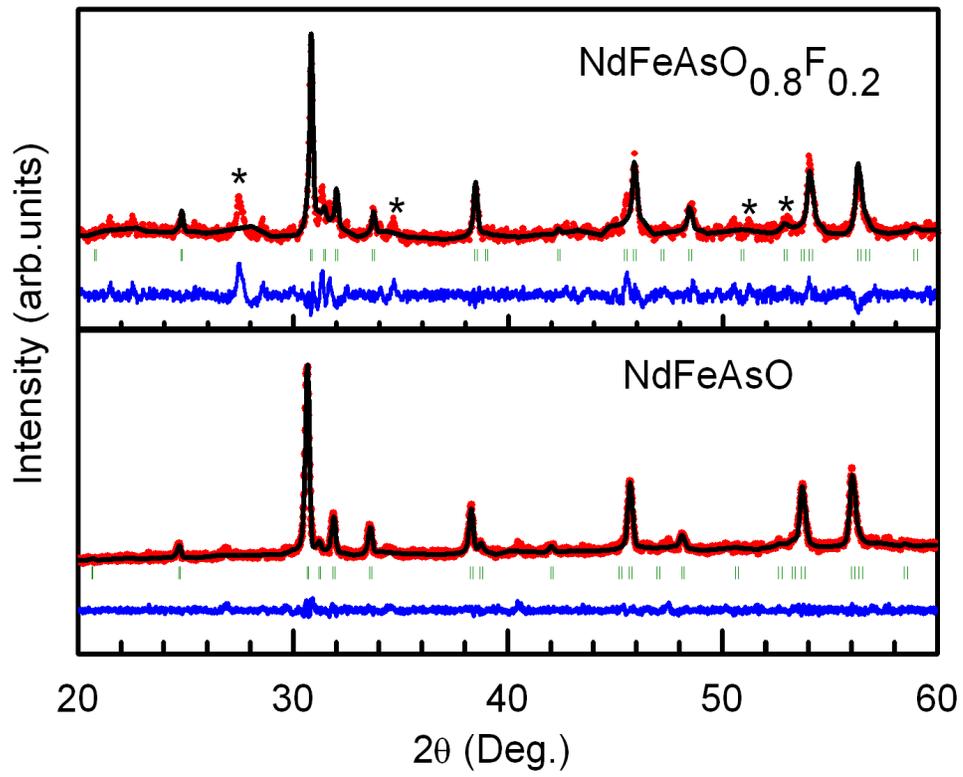

Figure 2 (a):

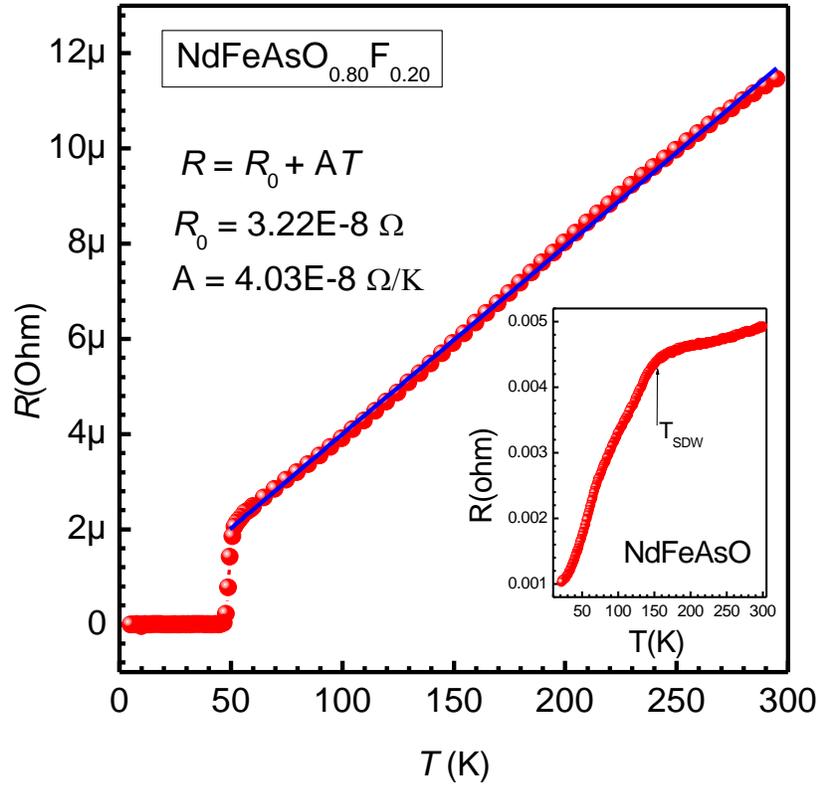

Figure 2 (b):

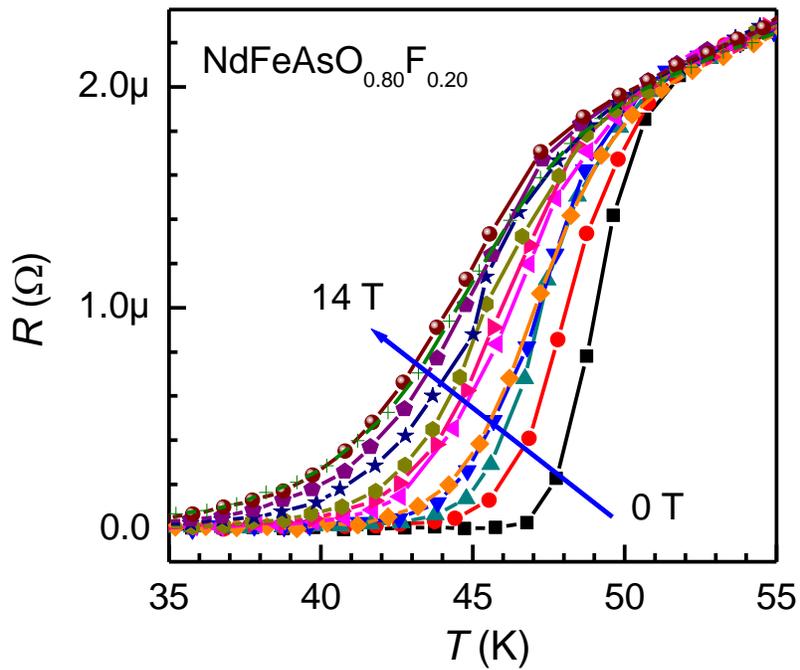

Figure 2 (c):

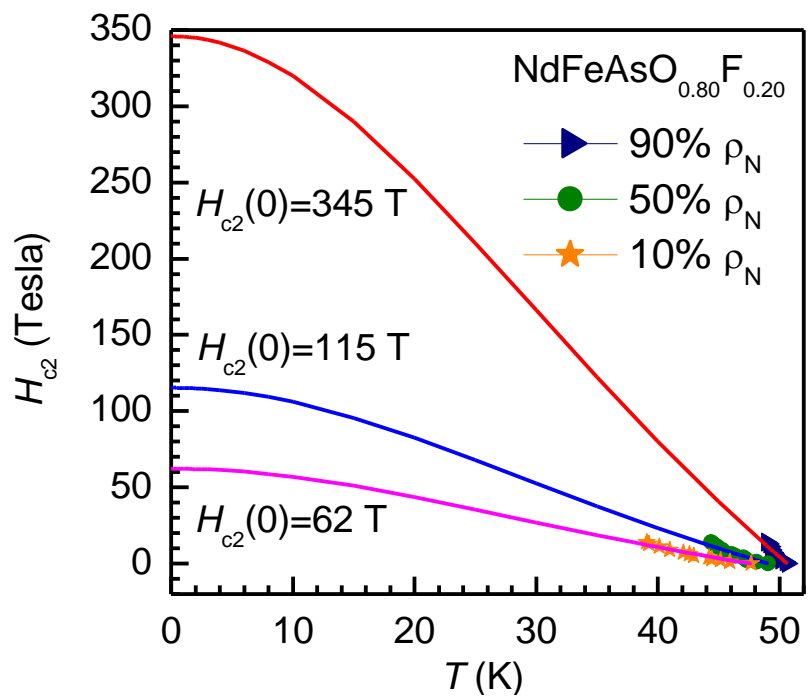

Figure 3:

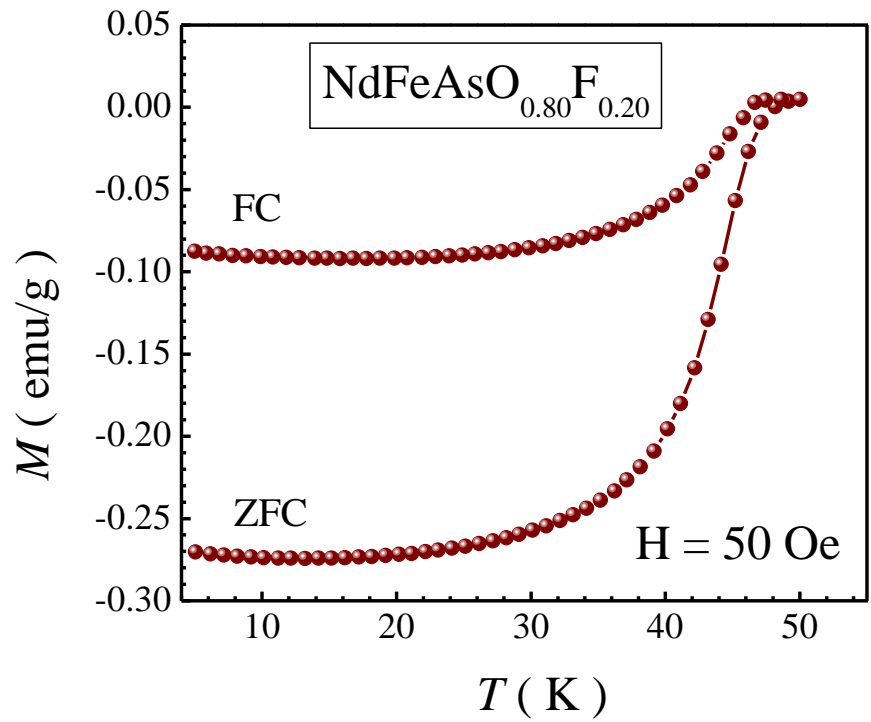

Figure 4:

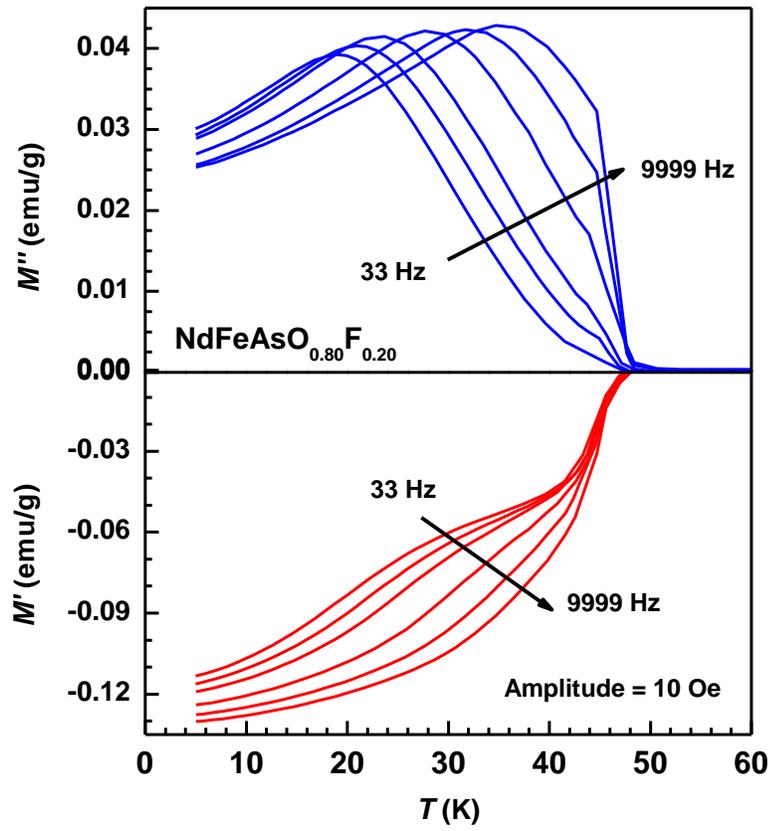

Figure 5:

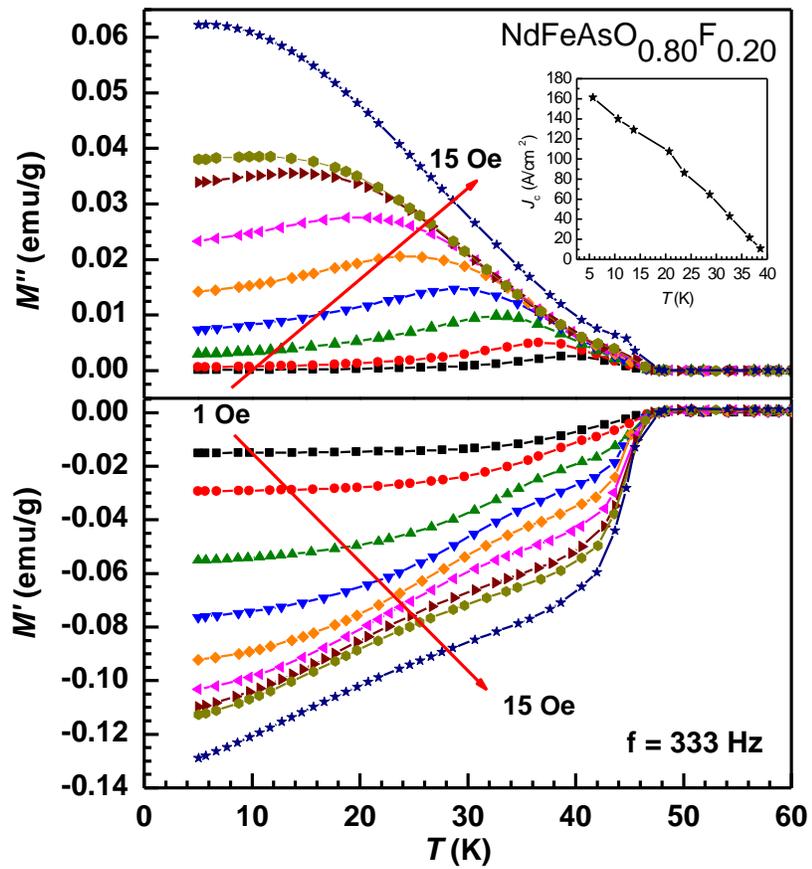

Figure 6:

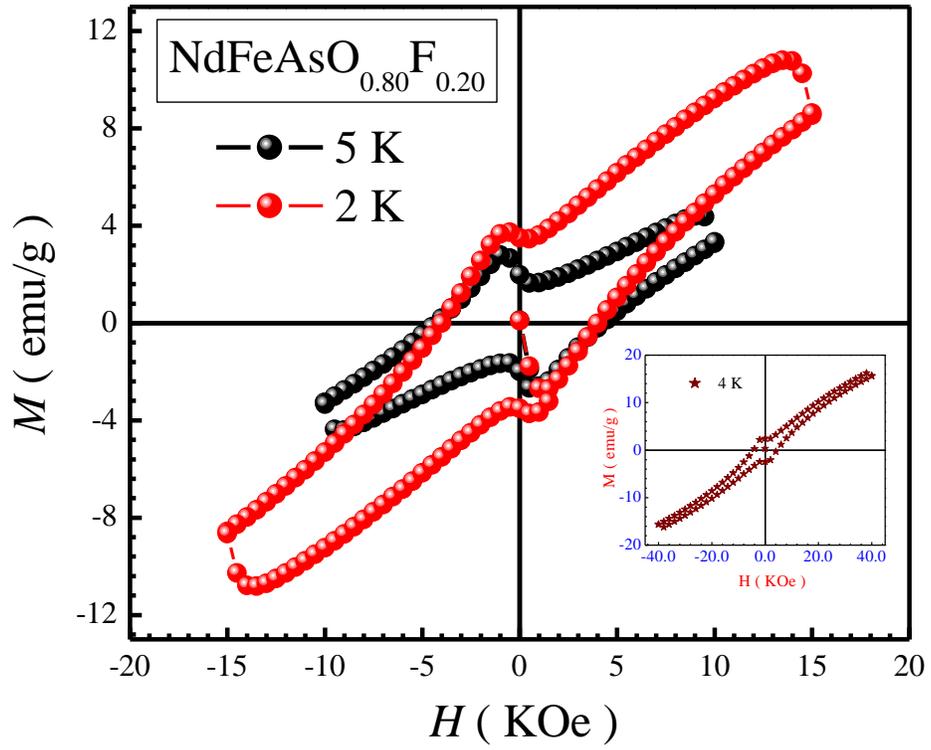

Figure 7:

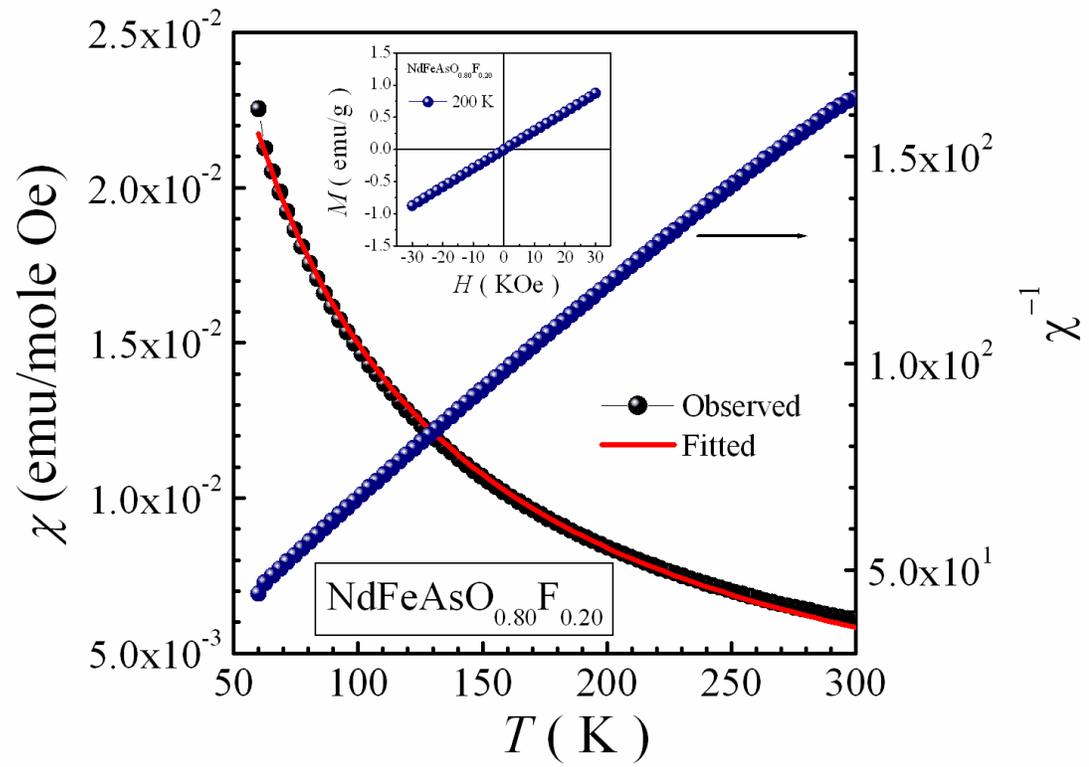

Figure 8:

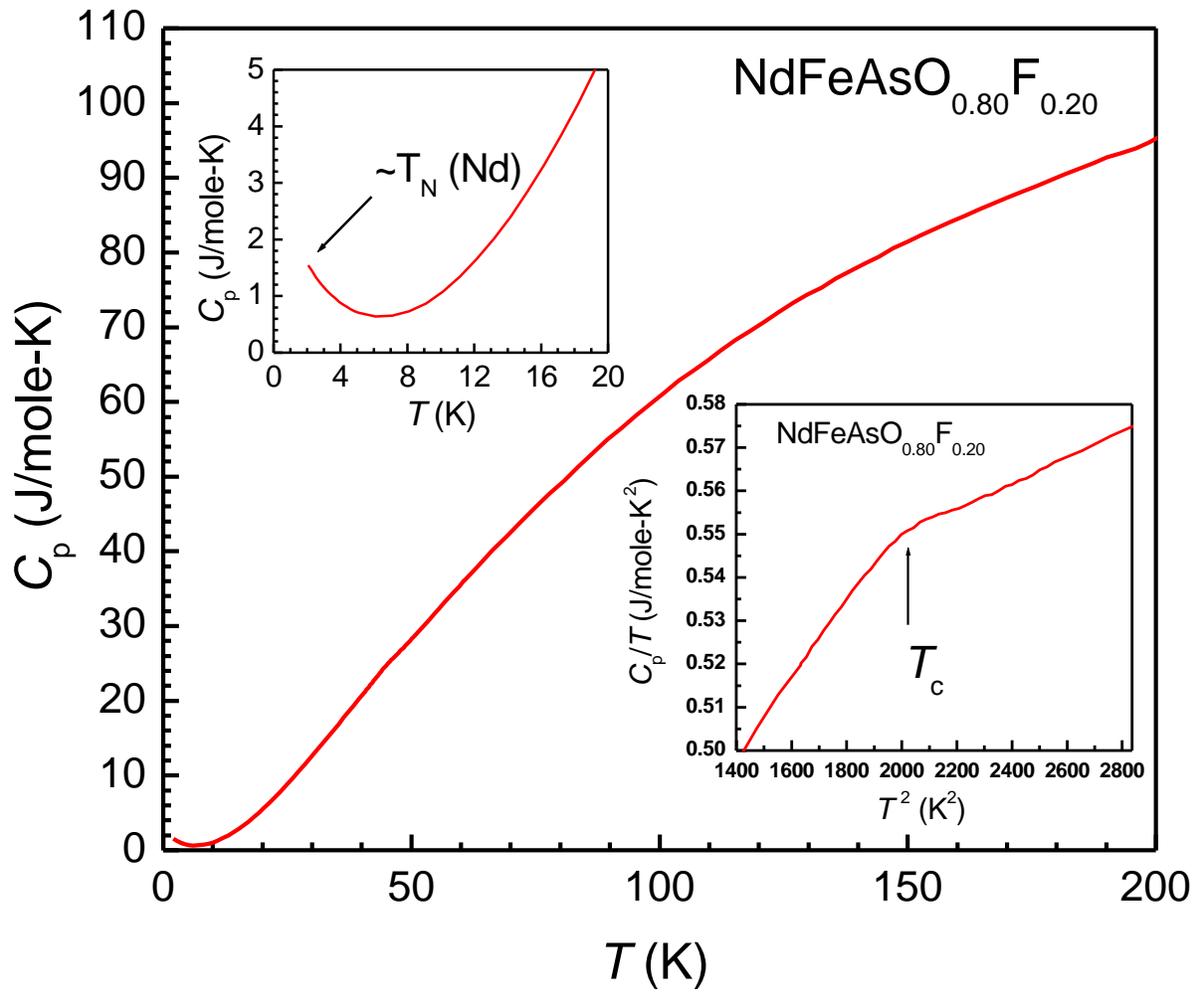

Figure 9:

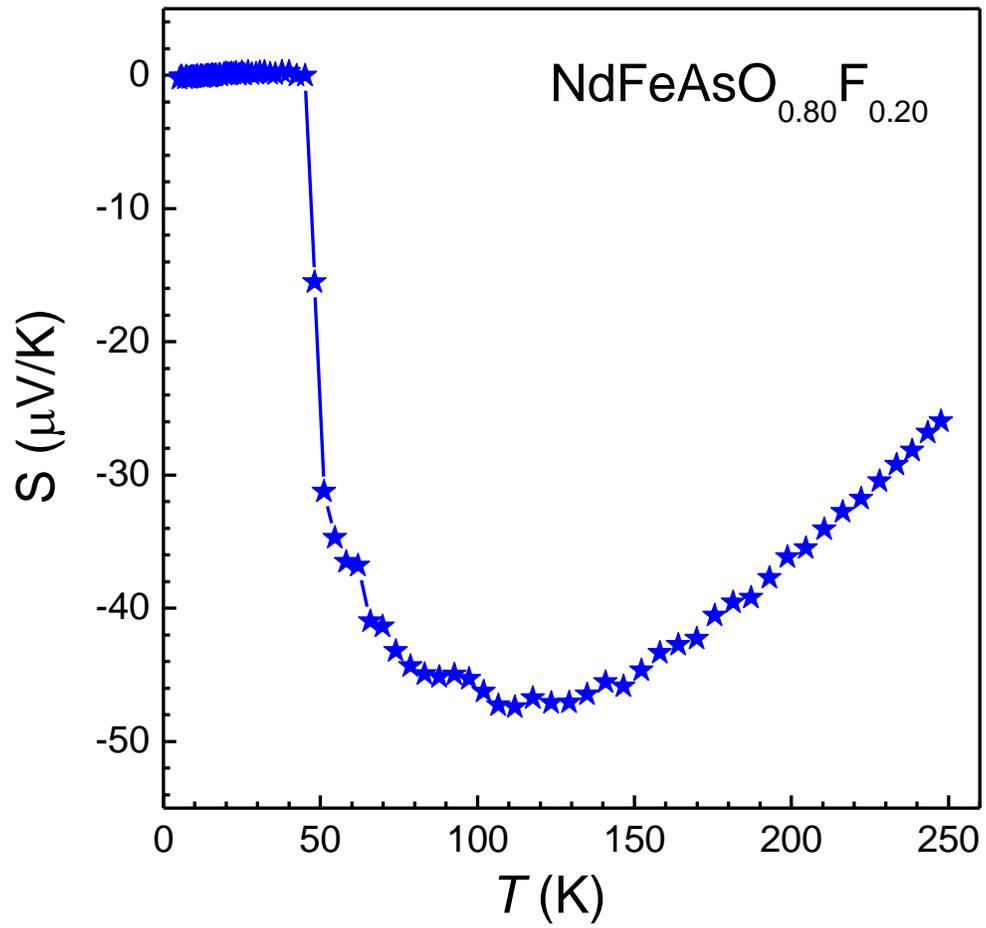